\documentclass{article}

\usepackage{arxiv}

\usepackage[utf8]{inputenc} 
\usepackage[T1]{fontenc}    
\usepackage{hyperref}       
\usepackage{url}            
\usepackage{booktabs}       
\usepackage{amsfonts}       
\usepackage{nicefrac}       
\usepackage{microtype}      
\usepackage{lipsum}    
\usepackage{graphicx}
\usepackage[square,numbers]{natbib}
\usepackage{doi}
\usepackage{xcolor}
\usepackage{upgreek}

\title{Thermophysical Properties of Spark Plasma Sintered UCo: a comparison with machine learning predictions}

\author{
  Yifan Sun\thanks{Correspondence to: sun.yifan.7r@kyoto-u.ac.jp}\\
  Kyoto University, Japan\\
  \And
  Hironobu Nakamura\\
  Osaka University, Japan\\
  \And
  Masaya Kumagai\thanks{Co-affiliated with SAKURA internet}\\
  Kyoto University, Japan\\
  \AND
  Yuji Ohishi\\
  Osaka University, Japan\\
  \And
  Ken Kurosaki\thanks{Correspondence to: kurosaki.ken.6n@kyoto-u.ac.jp}\\
  Kyoto University, Japan\\
}

\renewcommand{\headeright}{}
\renewcommand{\undertitle}{}
\renewcommand{\shorttitle}{}

\begin{document}
\maketitle

\begin{abstract}
  Uranium dioxide has been widely used as a nuclear fuel in commercial light water reactors due to its high uranium density and chemical stability. However, its relatively low thermal conductivity is not optimal from the viewpoints of fuel integrity and safety margins, particularly during loss-of-coolant accidents. Although the development of accident-tolerant fuels with higher thermal conductivity is strongly desired, many potential uranium compounds remain unexplored due to constraints associated with handling radioactive materials. To efficiently screen promising uranium compounds with high thermal conductivity, past studies have leveraged machine-learning models to accelerate the discovery process. In this study, we experimentally examine the model’s predictions by fabricating UCo and measuring its high-temperature thermophysical properties. Our results show that the thermal conductivity of UCo predicted by machine learning is in good agreement with the experimental measurements. Despite slight discrepancies, additional SHAP analysis suggests that the model’s decision logic is consistent with known physical trends. Overall, this study fills a gap in reported thermophysical properties of UCo and provides experimental support for machine-learning-assisted screening of uranium compounds relevant to advanced fuel development.
\end{abstract}


\section{Introduction}
The Fukushima Daiichi Nuclear Power Plant accident in 2011 renewed awareness of the need for accident-tolerant fuels (ATFs) that provide higher safety margins than the conventional UO\textsubscript{2}–Zircaloy fuel–cladding system in light water reactors (LWRs)~\cite{yun2021current,alrwashdeh2025critical}. Uranium dioxide has long been used as the standard LWR fuel because of its high melting point, chemical stability, and excellent irradiation tolerance~\cite{abrefah1994high,rest2019fission}. However, its low thermal conductivity leads to high fuel centerline temperatures and large temperature gradients, which can promote pellet cracking and enhance fission product release~\cite{pickman1972design}. Therefore, enhancing fuel thermal conductivity while preserving these favorable properties is an important direction in ATF development to improve fuel integrity and safety margins.

In the search for promising ATF candidates, uranium compounds such as UN~\cite{steven1988thermal}, UC~\cite{gokul2022uranium}, and U\textsubscript{3}Si\textsubscript{2}~\cite{white2015thermophysical,mohamad2018thermal} have been extensively investigated because they offer higher uranium density and thermal conductivity than UO\textsubscript{2}~\cite{ronchi1999thermal}. Higher uranium density is favorable for it improves fuel utilization and provides greater flexibility in core design and operating conditions~\cite{lewis2017fundamentals}. However, chemical stability issues limit their commercial deployment: UN~\cite{mikael2017uranium} and UC~\cite{bradley1962hydrolysis} react readily with water and steam, and U\textsubscript{3}Si\textsubscript{2}~\cite{migdisov2021instability} is also known to pulverize in high-temperature steam environments. While ongoing research has focused on improving the stability of these candidates~\cite{lopes2017degradation,mishchenko2021uranium,xu2025enhanced}, exploring the broader uranium compound space remains important for advanced fuel development.

For uranium compounds, where experimental investigations are limited by radiological and technical constraints, efficient material screening is necessary to guide targeted synthesis and discovery. In recent years, the availability of large materials databases~\cite{wilthan2017data,jain2013commentary,katsura2019data} has enabled machine learning (ML) models for materials property predictions~\cite{ward2016general,gaultois2016perspective,moses2021machine,sun2024multiclass}, offering a promising approach to efficiently identify ATF candidates. For example, in our previous work~\cite{sun2024multiclass}, we trained a classification model to screen uranium compounds with high thermal conductivity and obtained a curated list of 119 compounds predicted to exceed 15~W/mK at temperatures between 300 and 1000~K. However, because most of these uranium compounds have not yet been experimentally characterized, the validity of these predictions must be assessed systematically over time to evaluate the model's practical usefulness for ATF screening.

In this work, we selected uranium cobalt (UCo) from the 119 compounds predicted to possess high thermal conductivity for experimental validation. Although UCo is not intended as a practical LWR fuel material because of unfavorable neutron absorption (Co\textsuperscript{59}$\rightarrow$Co\textsuperscript{60}) and transmutation reactions (Co\textsuperscript{60}$\rightarrow$Ni\textsuperscript{60}), it is a suitable validation case because it has previously been successfully synthesized~\cite{baenziger1950compounds,chen1985superconductivity,takuya1991hydrogen,matsuda2011detailed}. However, prior studies have mainly focused on its low-temperature transport properties~\cite{chen1985superconductivity,matsuda2011detailed} or hydrogen absorption properties~\cite{takuya1991hydrogen}, and its high-temperature thermophysical properties have not yet been reported. Here, we synthesized near-single-phase UCo via arc melting and spark plasma sintering (SPS), measured its thermal conductivity, and compared the results with the ML predictions. Finally, to examine whether the model’s predictions are physically consistent, we performed SHapley Additive exPlanations (SHAP) analysis to interpret the model’s decision logic.

\section{Methodology}
\subsection{Synthesis and characterization}
The UCo bulk sample was synthesized by first weighing stoichiometric amounts of uranium and cobalt (99.9\% purity, Furuuchi Chemical Co., Ltd.) and then arc-melting the mixture under an argon atmosphere. The ingot was manually crushed in an argon-filled glove box to prevent oxidation. The resulting powder was loaded into a graphite die and densified by spark plasma sintering (SPS-515A, Sumitomo Coal Mining Co., Ltd.) to produce a sintered pellet. Spark plasma sintering was performed under a flowing argon atmosphere (200~mL/min) at a heating rate of 50~K/min up to 973~K, followed by a 1-hour dwell. The applied pressure was maintained at 50 MPa throughout the heating process and was increased to 100 MPa during the final 5~min of sintering.

The phase composition of the bulk UCo sample was assessed by X-ray diffraction (XRD; Rigaku Ultima IV) with Cu~K$\alpha$ radiation over a $2\uptheta$ range of 20--120$^\circ$. Lattice parameters were refined by least-squares fitting of the diffraction peak positions, using NIST silicon powder as an internal standard. Surface morphology was examined by scanning electron microscopy (SEM; JSM-6500F, JEOL), and elemental distribution was evaluated by energy-dispersive X-ray spectroscopy (EDS; EX-23000BU, JEOL).

\subsection{Property measurements}
The thermal diffusivity $\alpha$ of the UCo sample was measured by a laser flash apparatus (LFA-457, Netzsch) under a flowing argon atmosphere (200~mL/min) over a temperature range of 293--973~K. Three laser shots were collected at each measurement temperature to obtain an average $\alpha$. Thermal conductivity $\kappa$ was then calculated from $\alpha$, sample density $\rho$, and specific heat capacity $C_{\mathrm{p}}$ as $\kappa = \alpha C_{\mathrm{p}}\rho$. The sample density $\rho$ was measured at room temperature using the Archimedes method, and $C_{\mathrm{p}}$ was estimated based on the Neumann--Kopp rule. Here, the atomic heat capacities of U and Co were taken from the thermochemical database reported by Barin et al.~\cite{barin2013thermochemical}.

To further clarify the lattice ($\kappa_{\mathrm{lat}}$) and electronic ($\kappa_{\mathrm{ele}}$) contributions to the thermal conductivity of UCo, electrical conductivity $\sigma$ measurements were performed from 298--861~K under a flowing argon atmosphere (100~mL/min) with an in-house four-probe measurement system. This apparatus is identical to that used in our previous study~\cite{sun2024thermophysical}, and its measurement accuracy has been validated using nickel as a standard reference material. $\kappa_{\mathrm{ele}}$ was calculated from the measured $\sigma$ using the Wiedemann--Franz law, $\kappa_{\mathrm{ele}} = L\sigma T$. The Lorenz number $L$ was taken as $2.44\times 10^{-8}\,\mathrm{W\,\Omega\,K^{-2}}$, which is the Sommerfeld value for metals. The lattice thermal conductivity $\kappa_{\mathrm{lat}}$ was obtained by subtracting $\kappa_{\mathrm{ele}}$ from the total thermal conductivity $\kappa$.

Vickers hardness measurements of the UCo sample were performed using a micro-Vickers hardness tester (HMV-G305, Shimadzu) with an applied load of 9.807~N and a dwell time of 10~s. Five consecutive indentations were made, and the average value was reported. For each indentation, the Vickers hardness $H_{\mathrm{V}}$ (kgf/mm$^{2}$) was calculated as $H_{\mathrm{V}} = 0.1891\,F/d^{2}$, where $F$ is the applied load (N) and $d$ is the diagonal length of the indentation (mm), and then converted to SI units (GPa). All Vickers hardness measurements were carried out at a room temperature of 25$^\circ$C.

\subsection{Machine learning and SHAP analysis}
In our previous work~\cite{sun2024multiclass}, we trained a random forest classifier to predict whether the thermal conductivity of a given uranium compound falls into three classes ($\leq 5$, $5<\kappa\leq 15$, and $>15$~W/mK). The model used 168,918 composition--temperature--thermal conductivity records from Starrydata~\cite{katsura2019data} and 133 input features (132 Magpie descriptors~\cite{ward2016general} plus temperature).

Because Starrydata has been continuously updated since our initial study (based on the 2023-01-12 release), we retrained the model using the same architecture and hyperparameters on the latest Starrydata release (2026-01-01) after basic filtering. Specifically, the temperature and thermal conductivity ranges of the training data were restricted to 300--1000~K and 0.001--500~W/mK, respectively, to remove outliers and retain only context-relevant data. Invalid chemical formulas were removed before generating the 132 Magpie descriptors~\cite{ward2016general}. During training, we used the same 133 input features (temperature and Magpie descriptors) to predict the same three thermal-conductivity classes. The Synthetic Minority Over-sampling Technique (SMOTE)~\cite{chawla2002smote} was applied to balance the number of data points in each class when training the random forest classifier with default hyperparameters. The updated model, trained using 223,023 data points, showed no substantial change in performance, which we attribute to the fact that most of the training data were already included in the original training set~\cite{sun2024multiclass}. This updated model was then used to predict the thermal conductivity of UCo from 300--1000~K. 

Additionally, to better understand the model's decision logic when predicting the thermal conductivity of UCo, SHAP analysis was performed with feature perturbation set to "tree path dependent" without an additional external background dataset. Compared with conventional feature-importance outputs, which provide an overview of globally important features during model training, SHAP offers a more direct interpretation of how much each input feature contributed to the final prediction for the thermal conductivity of UCo at a given temperature, allowing us to qualitatively assess whether the prediction is consistent with known materials-science principles.

\section{Results}
\subsection{Sample characterizations}
\begin{figure}[h]
  \centering
  \includegraphics[width=0.48\linewidth]{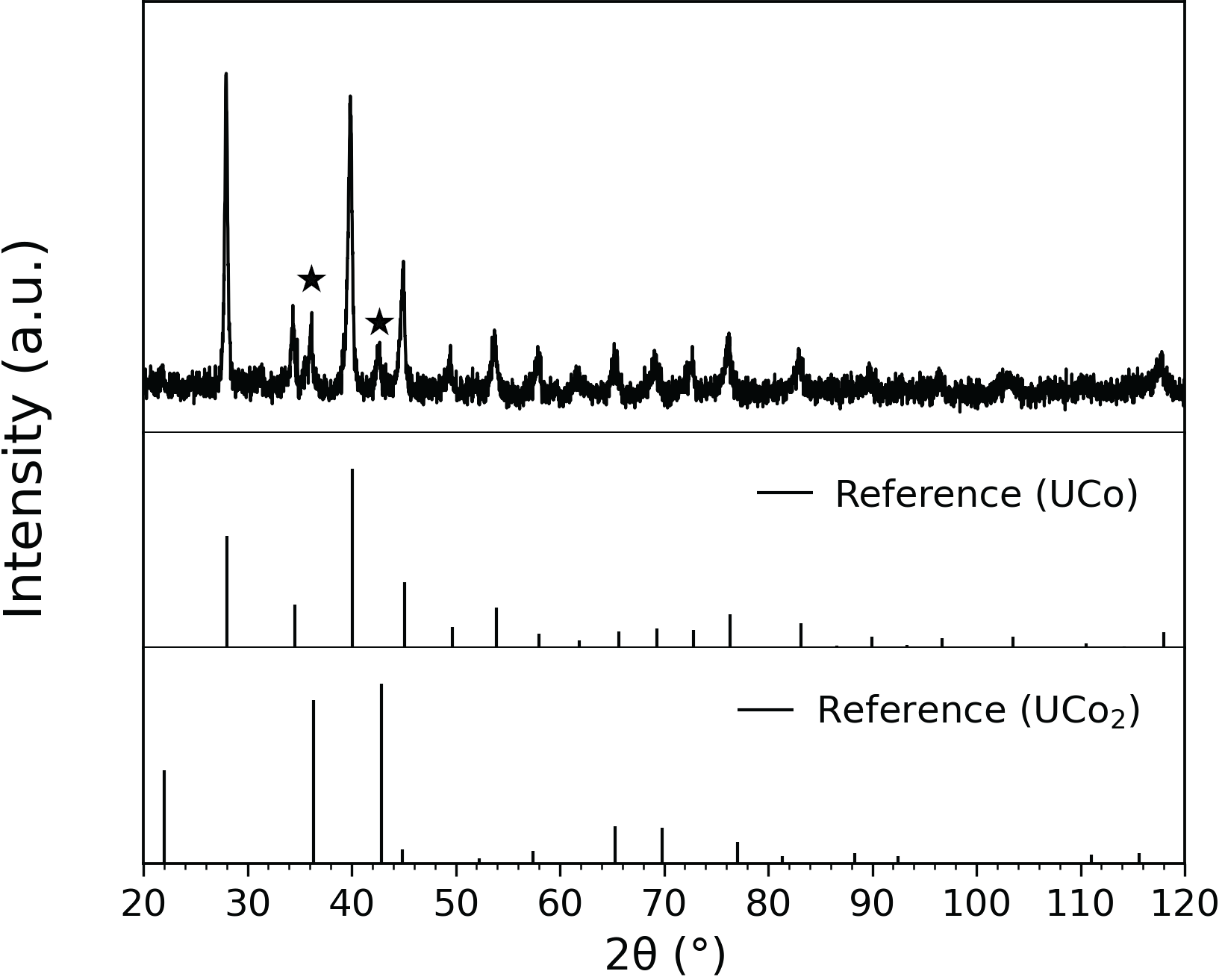}
  \caption{XRD pattern of the prepared UCo sample after SPS, along with reference patterns for UCo~\cite{baenziger1950compounds} and UCo\textsubscript{2}~\cite{smith1975icdd}.}
  \label{fig:xrd_pattern}
\end{figure}

Figure~\ref{fig:xrd_pattern} shows the XRD pattern of the prepared UCo sample after SPS. Most diffraction peaks can be assigned to the cubic UCo phase~\cite{baenziger1950compounds}, indicating that the bulk sample is predominantly single-phase UCo. Two weak reflections observed near 36.3$^\circ$ and 42.9$^\circ$ (marked by stars in Fig.~\ref{fig:xrd_pattern}) were assigned to UCo\textsubscript{2}~\cite{smith1975icdd}, suggesting the presence of a minor secondary phase. The refined lattice parameter of cubic UCo shows good agreement with previously reported values~\cite{baenziger1950compounds}, as summarized in Table~\ref{tab:UCo_lattice_density}. Although calculating the relative density (TD\%) is not straightforward due to the presence of a minor UCo\textsubscript{2} phase, the measured density (14.6 g/cm\textsuperscript{3}) being close to the theoretical density of UCo (15.1 g/cm\textsuperscript{3}) still indicates that the SPS process yielded a highly densified sample.

\begin{table}[h]
  \centering
  \caption{Lattice parameter and density of the synthesized UCo sample}
  \label{tab:UCo_lattice_density}
  \begin{tabular}{lccc}
    \toprule
    &
    \begin{tabular}{c}
      Lattice parameter\\
      (nm)
    \end{tabular}
    &
    \begin{tabular}{c}
      Theoretical density\\
      (g/cm$^{3}$)
    \end{tabular}
    &
    \begin{tabular}{c}
      Measured density\\
      (g/cm$^{3}$)
    \end{tabular}
    \\
    \midrule
    This work      & 0.6389 & 15.1 & 14.6 \\
    Reference~\cite{baenziger1950compounds}& 0.6356 & 15.4 & --- \\
    \bottomrule
  \end{tabular}
\end{table}

\begin{figure}[h]
  \centering
  \includegraphics[width=0.95\linewidth]{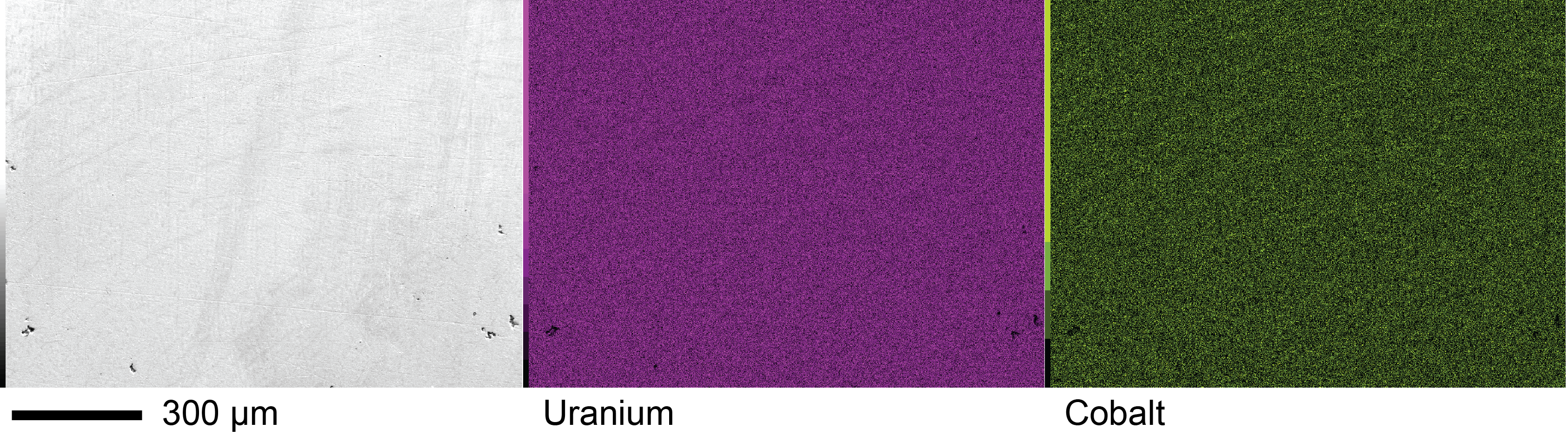}
  \caption{SEM-EDS images of the prepared UCo sample after SPS.}
  \label{fig:UCo_SEM_image}
\end{figure}

The SEM image in Figure~\ref{fig:UCo_SEM_image} shows a dense microstructure with no significant porosity, as reflected in the sample's high relative density. EDS mappings reveal that U and Co are homogeneously distributed on a larger scale, consistent with the XRD results. The formation of UCo\textsubscript{2} impurities has also been reported in a previous study using arc melting~\cite{takuya1991hydrogen}. Based on the U--Co phase diagram~\cite{okamoto2017supplemental}, UCo does not melt congruently and, upon cooling, is produced via the peritectic reaction between the liquid phase and UCo\textsubscript{2}, which likely contributes to the difficulty of synthesizing single-phase UCo. Therefore, based on the XRD peak intensities and EDS mapping, we regard the our prepared UCo sample as near-single-phase with minor UCo\textsubscript{2} impurities.

\subsection{Thermophysical transport properties}
\begin{figure}[h]
  \centering
  \includegraphics[width=0.95\linewidth]{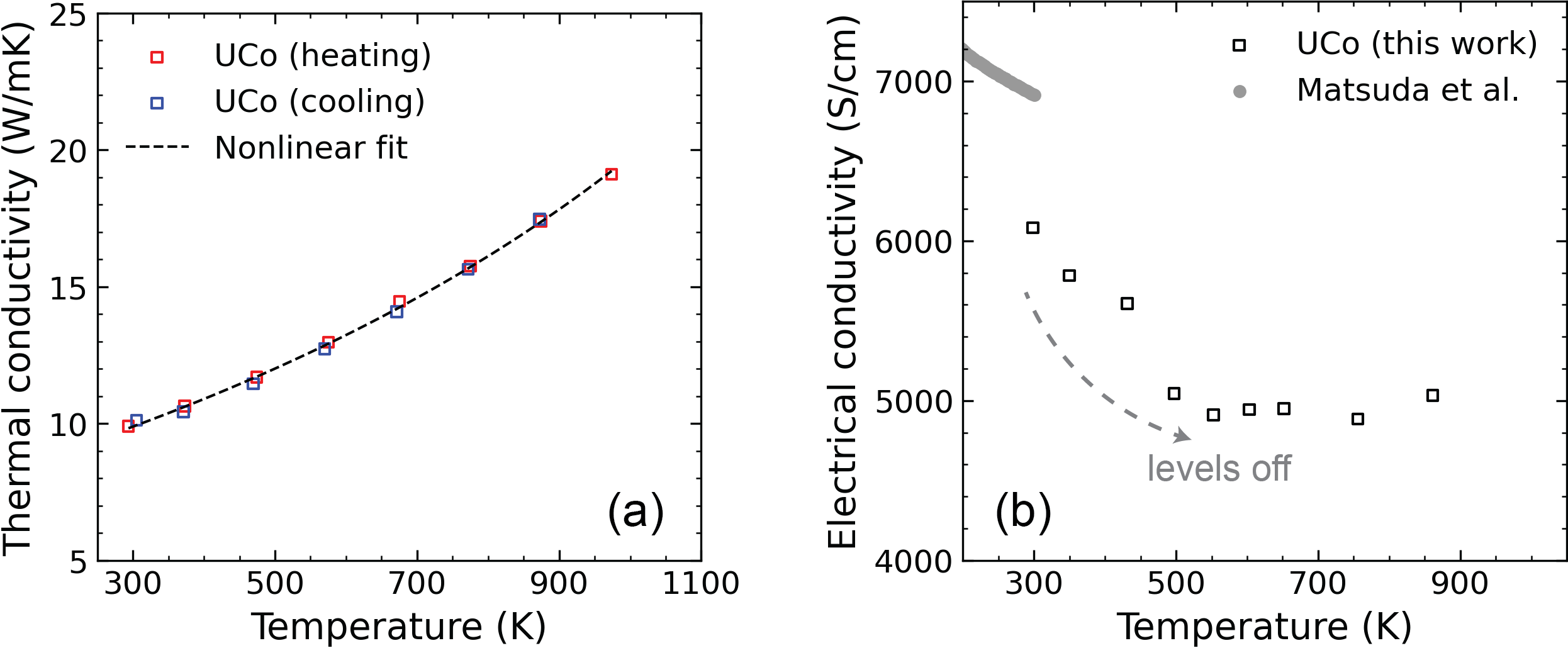}
  \caption{(a) Thermal conductivity and (b) electrical conductivity of the prepared UCo bulk sample. The arrow in (b) highlights that the electrical conductivity approaches a plateau as the temperature increases above 500~K.}
  \label{fig:UCo_transport}
\end{figure}

The measured thermal conductivity and electrical conductivity of UCo are summarized in Figure~\ref{fig:UCo_transport}. Consistent with its metallic nature, UCo exhibits a moderate thermal conductivity of 9.9~W/mK at 294~K, which increases monotonically with temperature and reaches 19.1~W/mK at 974~K. Unlike uranium borides~\cite{kardoulaki2020thermophysical}, no hysteresis was observed between the heating and cooling runs. Within the measurement temperature range (294--974~K), the temperature dependence can be well described by the following equation, which has been used to fit the thermal conductivity of semimetallic UB$_4$~\cite{kardoulaki2020thermophysical}:

\begin{equation}
  \label{eq:total_kappa}
  \kappa = \frac{T}{A T + B} + C,
\end{equation}
where $A = -5.326\times 10^{-2}\,\mathrm{mK/W}$, $B = 1.338\times 10^{2}\,\mathrm{mK^{2}/W}$, and $C = 7.343\,\mathrm{W/mK}$.

Figure~\ref{fig:UCo_transport}(b) shows the electrical conductivity of UCo measured using our in-house four-probe apparatus under Ar flow. At around 300~K, the measured conductivity is slightly lower than that reported by Matsuda et al.~\cite{matsuda2011detailed} for polycrystalline UCo. This deviation may result from differences in impurity phases between the samples. Unlike the UCo\textsubscript{2} secondary phase observed in this work, U$_6$Co impurities were present in the sample synthesized by Matsuda et al.~\cite{matsuda2011detailed}. With increasing temperature, UCo initially exhibits the expected metallic behavior, with electrical conductivity decreasing approximately linearly with temperature. However, above 500~K, the decrease in electrical conductivity levels off and approaches a plateau. We interpreted this as resistivity saturation, where the electron mean free path becomes comparable to the interatomic spacing, limiting further increases in electrical resistivity from phonon-electron scattering with temperature.

\begin{figure}[h]
  \centering
  \includegraphics[width=0.48\linewidth]{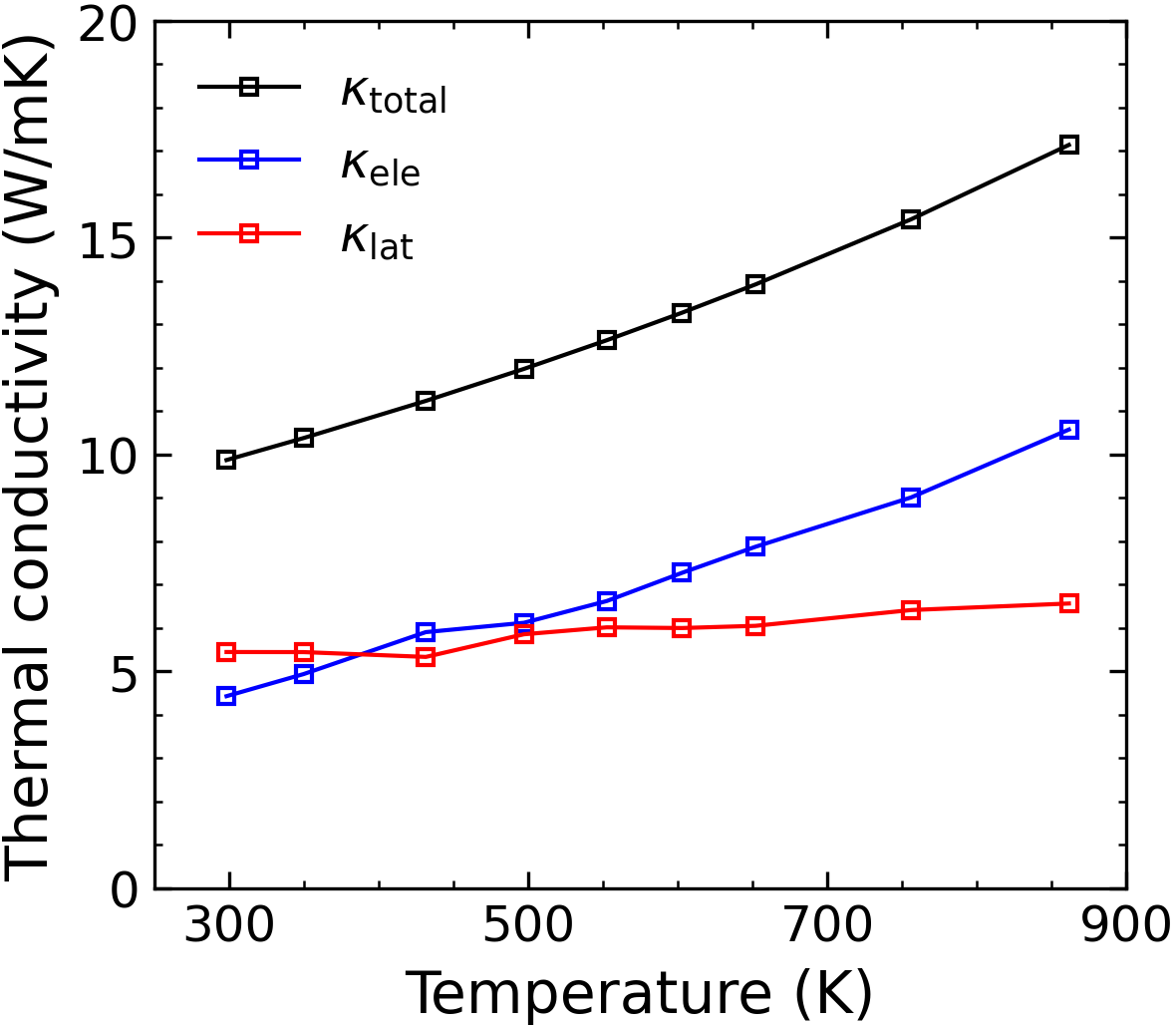}
  \caption{Temperature dependence of thermal conductivity $\kappa_{\mathrm{total}}$, electronic thermal conductivity $\kappa_{\mathrm{ele}}$, and lattice thermal conductivity $\kappa_{\mathrm{lat}}$ of the prepared UCo sample.}
  \label{fig:kappa_components}
\end{figure}

Based on the Wiedemann--Franz law, the relatively constant electrical conductivity of UCo implies that the electronic thermal conductivity increases approximately linearly with temperature. At elevated temperatures where lattice thermal conductivity is limited by phonon--phonon Umklapp scattering, electronic contribution is expected to dominate the total thermal conductivity. This is illustrated in Figure~\ref{fig:kappa_components}, where the lattice thermal conductivity only increases slightly from 5.4 to 6.6~W/mK as the temperature increases from 298 to 861~K, whereas the electronic thermal conductivity increases from 4.4 to 10.6~W/mK. Nevertheless, phonon transport still plays a non-negligible role up to approximately 600~K, where the electronic and lattice contributions are comparable. Finally, the slight apparent increase in the lattice thermal conductivity with temperature in Figure~\ref{fig:kappa_components} is considered to result from using a constant Lorenz number when estimating the electronic contribution.

\subsection{Vickers Hardness Test}
\begin{figure}[h]
  \centering
  \includegraphics[width=0.48\linewidth]{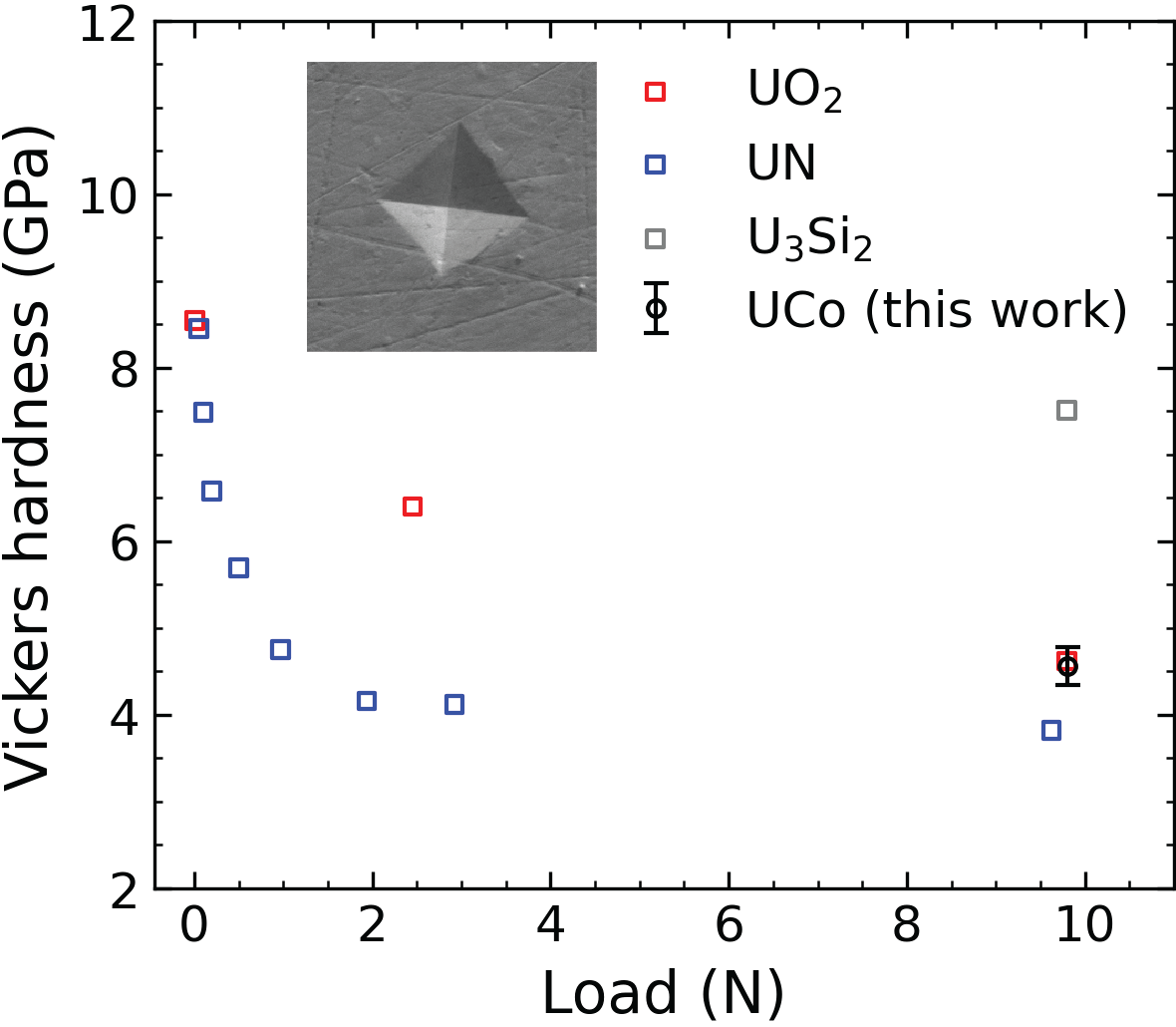}
  \caption{Comparison of the Vickers hardness of UCo with that of other uranium compounds at various indentation loads. The inset shows a Vickers indentation on the UCo sample under a load of 9.807~N.}
  \label{fig:vickers_hardness}
\end{figure}

An example of a Vickers hardness indentation is shown in the inset of Figure~\ref{fig:vickers_hardness}. Under a load of 9.807~N, the Vickers hardness of UCo was determined to be 4.6~GPa at room temperature and is compared with reported data for U\textsubscript{3}Si\textsubscript{2}~\cite{mohamad2018thermal}, UO\textsubscript{2}~\cite{yamada1998mechanical,kurosaki2004nanoindentation}, and UN~\cite{adachi2009mechanical}. Overall, the Vickers hardness of intermetallic UCo is significantly lower than that of U\textsubscript{3}Si\textsubscript{2} and comparable to that of uranium-based ceramics such as UO\textsubscript{2} and UN.

\section{Discussion}
\begin{figure}[h]
  \centering
  \includegraphics[width=0.48\linewidth]{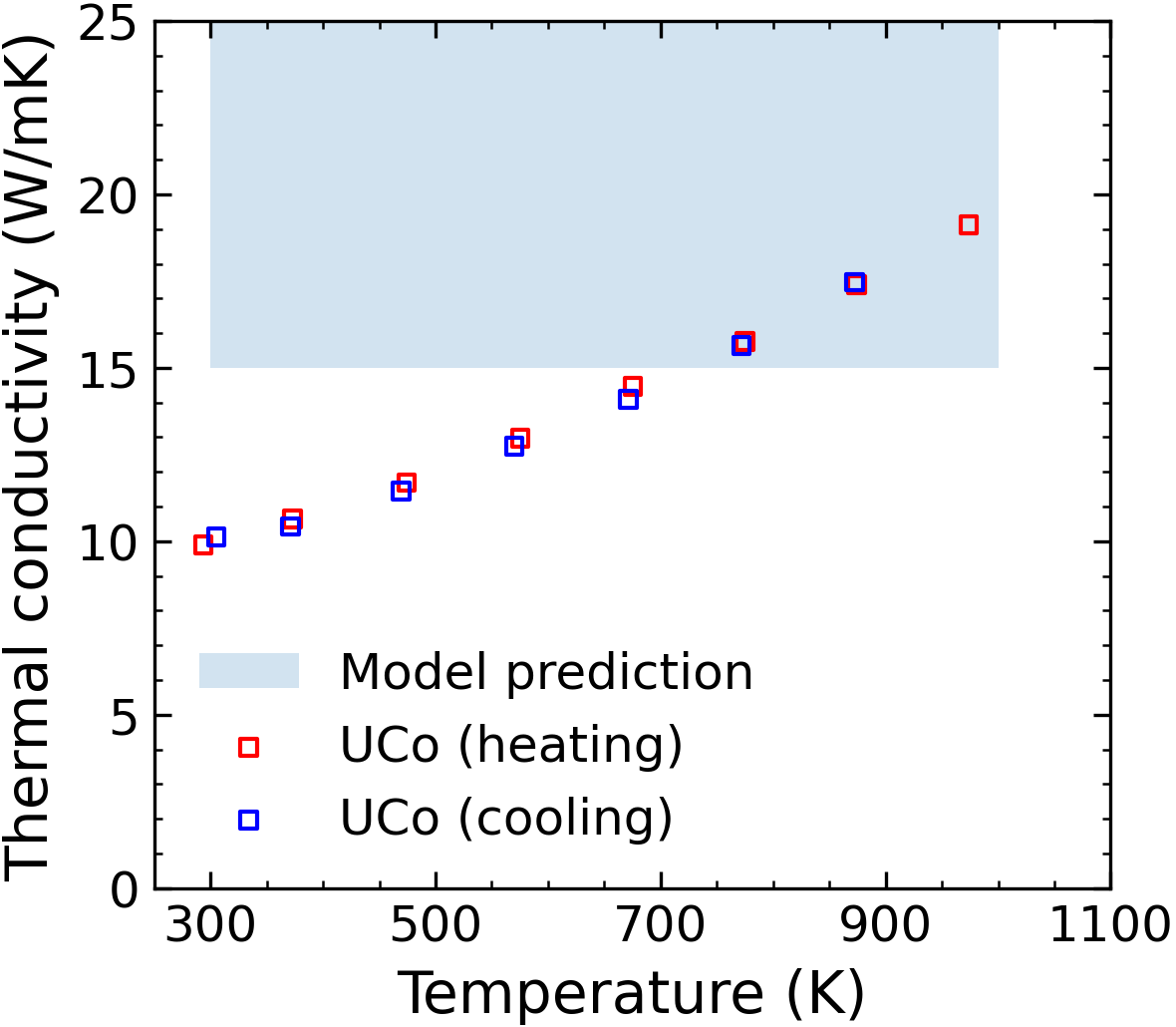}
  \caption{Comparison between the experimentally determined thermal conductivity of UCo and the machine-learning model prediction. The model's predicted region ($\kappa>15$~W/mK) is shaded in blue.}
  \label{fig:model_comparison}
\end{figure}

The experimentally determined thermal conductivity of UCo is compared with our classification model's predicted range from 300 to 1000~K in Figure~\ref{fig:model_comparison}, and the model output probabilities for each "class" are summarized in Table~\ref{tab:model_class_probabilities}. Over the entire temperature range, the model predicts that UCo exhibits a thermal conductivity greater than 15~W/mK. Compared with the experimental values, the model constantly overestimates $\kappa$ below 700~K, indicating that its quantitative accuracy has room for improvement. Nevertheless, the probabilities in Table~\ref{tab:model_class_probabilities} suggest that the model is not making arbitrary predictions. In particular, it consistently assigns a low probability to the $\kappa\leq 5$~W/mK range, correctly indicating that UCo should not have a very low thermal conductivity based on its chemical features. Instead, at lower temperatures, the model appears to mainly struggle to distinguish between the $5<\kappa\leq 15$~W/mK and $\kappa>15$~W/mK ranges, i.e., whether UCo is a ``good'' or ``excellent'' thermal conductor. This struggle is consistent with physical intuition: while metals often exhibit relatively high thermal conductivity at elevated temperatures due to significant electronic contribution, the magnitude at lower temperatures can be more difficult to infer from composition-based descriptors alone. Overall, this behavior suggests that even though the thermal conductivity class boundary is not always predicted correctly, the model can qualitatively capture UCo's relatively high thermal conductivity.

\begin{table}[h]
  \centering
  \caption{Machine-learning model predictions for the thermal conductivity of UCo between 300 and 1000~K. Thermal conductivity is reported in W/mK.}
  \label{tab:model_class_probabilities}
  \begin{tabular}{ccccc}
    \toprule
    Temperature (K) & Predicted range & $P(\kappa\leq 5)$ & $P(5<\kappa\leq 15)$ & $P(\kappa>15)$ \\
    \midrule
    300 & $\kappa>15$ & 0.09 & 0.41 & 0.50 \\
    400 & $\kappa>15$ & 0.08 & 0.40 & 0.52 \\
    500 & $\kappa>15$ & 0.09 & 0.41 & 0.50 \\
    600 & $\kappa>15$ & 0.09 & 0.43 & 0.48 \\
    700 & $\kappa>15$ & 0.09 & 0.38 & 0.53 \\
    800 & $\kappa>15$ & 0.09 & 0.35 & 0.56 \\
    900 & $\kappa>15$ & 0.09 & 0.34 & 0.57 \\
    1000 & $\kappa>15$ & 0.09 & 0.34 & 0.57 \\
    \bottomrule
  \end{tabular}
\end{table}

\begin{figure}[h]
  \centering
  \includegraphics[width=0.95\linewidth]{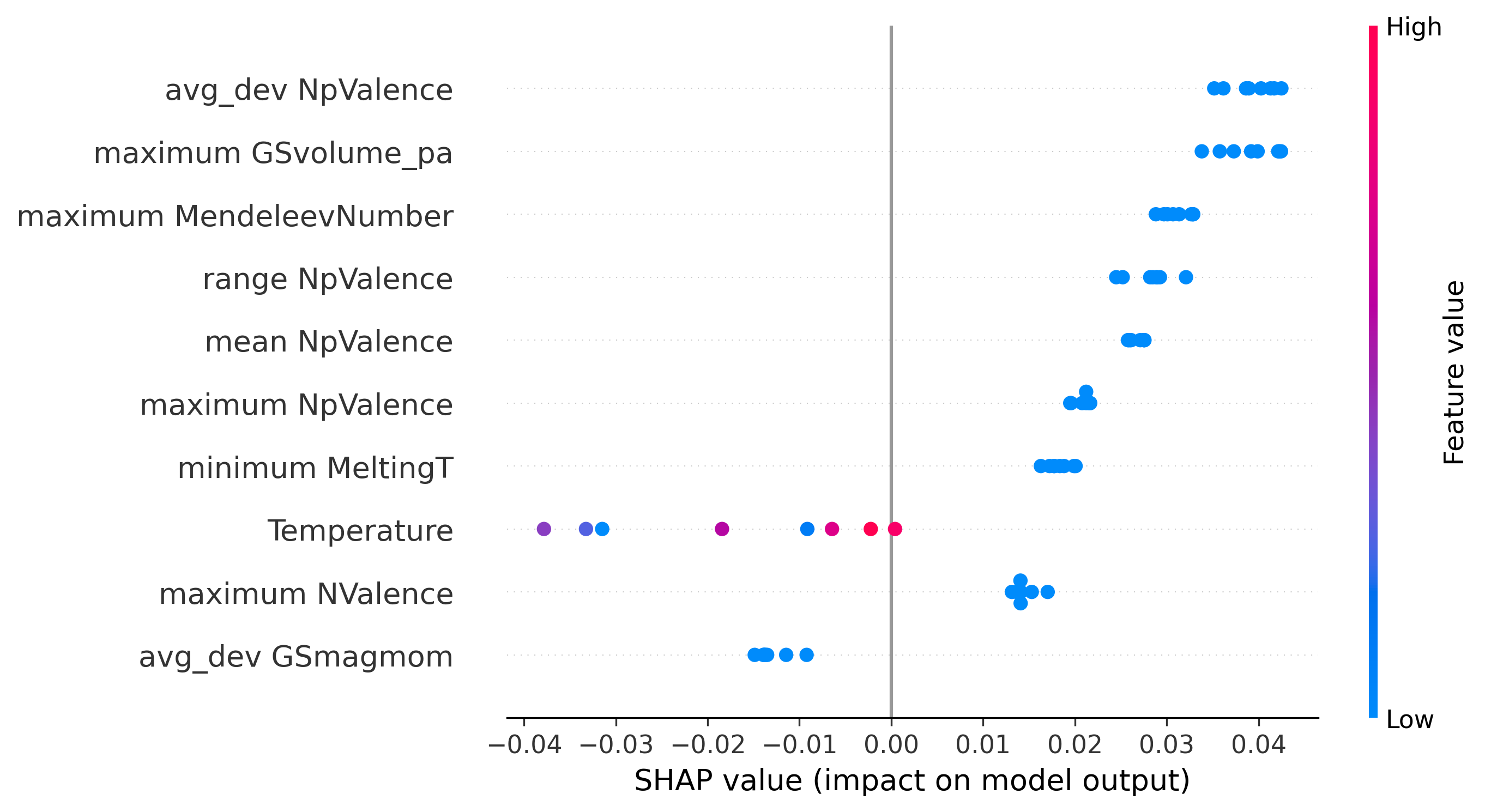}
  \caption{SHAP beeswarm plot showing the top 10 features that contribute most to the model classifying UCo as having $\kappa>15$~W/mK.}
  \label{fig:shap_beeswarm}
\end{figure}

Follow-up SHAP analysis was then performed to identify which input chemical features most strongly supported the model classifying UCo as a high-thermal-conductivity compound ($\kappa>15$~W/mK). These results are important for assessing whether the model's decision logic is qualitatively consistent with physical expectations. The SHAP values of the 10 most important features supporting the $\kappa>15$~W/mK class prediction for UCo evaluated from 300 to 1000~K are shown in the beeswarm plot in Figure~\ref{fig:shap_beeswarm}. Here, only the temperature feature has a color gradient because it varied from 300 to 1000~K, whereas Magpie features are constant for a fixed composition (e.g., UCo).

In Figure~\ref{fig:shap_beeswarm}, among the features with the highest SHAP value are several related to p-valence electron counts (NpValence). Within the Magpie feature set, low NpValence statistics can serve as a rough compositional fingerprint for metallic or intermetallic compounds that are dominated by transition metals with p-valence electron counts of 0. Simultaneously, such systems often exhibit high thermal conductivity because of a substantial electronic contribution to heat transport. In the case of UCo, the NpValence-related Magpie features (average deviation, range, mean, and maximum) are also all 0. Therefore, the prominence of these NpValence features in Figure~\ref{fig:shap_beeswarm} suggests that the model is using physically meaningful proxies for metallic characteristics to support its high-thermal-conductivity ($\kappa>15$~W/mK) prediction for UCo. In addition, Figure~\ref{fig:shap_beeswarm} shows the SHAP value of temperature becomes more positive with increasing temperature, indicating that higher temperature pushes the model toward selecting the $\kappa>15$~W/mK class, as reflected in the probability outputs in Table~\ref{tab:model_class_probabilities}. This trend is also qualitatively consistent with observations that thermal conductivity in metallic system often increases with elevating temperature due to its significant electronic contributions.

Overall, the SHAP results suggest that our trained classification model relies on physically meaningful indicators (NpValence-related features) and temperature dependence when predicting the thermal conductivity class of UCo from 300 to 1000~K, rather than relying on arbitrary inputs. However, intermetallics such as UCo are chemically simple and may represent a relatively easy case for the model to identify using valence-electron descriptors. Therefore, for a more stringent evaluation of model performance, chemically more complex uranium compounds, as well as outlier intermetallics predicted to have low thermal conductivity, will be synthesized and measured in future validation studies.

\section{Conclusion}
In this study, we prepared a dense, near-single-phase UCo pellet by spark plasma sintering and measured its thermal conductivity to assess the performance of our previously proposed machine learning classification model. The thermal conductivity of UCo increases monotonically with temperature, from 9.9~W/mK at 294~K to reaching 19.1~W/mK at 974~K.

Comparison with the model predictions shows that this classifier constantly overestimates the thermal conductivity of UCo below $\sim$700~K. From the model output probabilities, we find that while the model reliably distinguishes UCo from the low-thermal-conductivity class ($\kappa\leq5$~W/mK), it is substantially less confident near the $\kappa=15$~W/mK threshold, contributing to the observed overestimation. SHAP analysis reveals that the model's decision logic is qualitatively physics-consistent, utilizing NpValence-related input features to capture UCo's intermetallic character and guide its prediction of $\kappa>15$~W/mK over 300--1000~K. Consequently, despite quantitative discrepancies at lower temperature, the model still provides a well-grounded prediction for the thermal conductivity of UCo. Finally, for a more stringent evaluation of model robustness in screening uranium compounds, we will extend future validation to chemically more complex uranium systems.

\section*{Acknowledgment}
This work was supported in part by MEXT Innovative Nuclear Research and Development Program, grant number 20354330 \& 22682541.


\bibliographystyle{unsrtnat}
\bibliography{references}  






\end{document}